# An Approach for Selecting Cloud Service Adequate to Big Data
## Case Study: E-health Context

Fatima Ezzahra MDARBI [1], Nadia AFIFI [1], Imane HILAL[1,2],Hicham BELHADAOUI [1]

[1] RITM Lab, EST, CED ENSEM
University Hassan II
Casablanca, Morocco

[2] Lyrica labs
Information Science School
Rabat, Morocco

Fati.mdarbi@gmail.com, Nafifi@est-uh2c.ac.ma, Ihilal@esi.com, belhadaoui_hicham@yahoo.fr

*Abstract*-The expanding Cloud computing's services offers great opportunities for consumers to find the best service and best cost. It offers a computing power and a storage space adapted especially for Big Data processing. However, it raises new challenges on how to select the best service out of the huge pool. It is time-consuming for consumers to collect the necessary information and analyze all service providers to make the right decision. Moreover, it' is a highly demanding task from a computational perspective, because the same computations may be conducted repeatedly by multiple consumers who have similar requirements. Therefore, in this paper, we propose an approach based on Analytic Hierarchy Process (AHP) method, which manages the selection of the Cloud Service adequate to Big Data based on its parameters and criteria. We applied this approach on a case study in order to validate its efficity. The studied case is about the selection of the adequate Cloud Service for Big Data in the context of National Health Service (NHS) of United Kingdom (UK).

## I. INTRODUCTION

Organizations are unable to manage, manipulate, process, share, retrieve, and analyze the Big Data using traditional software tools. Those latter are generally costly and time-consuming during data processing. Big Data is a collection of complex data with massive volume; it needs tools for a real time data management and analysis capabilities. Whereas Cloud computing is the distributed computing model, it is a trending solution for analyzing Big Data, and providing both computing facilities and resources for users. The aim of the Cloud model is to increase the opportunities for Cloud users by accessing leased infrastructure and software applications from anywhere anytime and for any data. However, the strong growth of Cloud Services makes it difficult for potential users to decide which options are best suited to their needs. Companies need to take a rational approach to ensure they choose the most appropriate Cloud Service for their Big Data.

Many approaches have dealt with the problem of Cloud Service selection. Among these approaches, we identify AHP. AHP is a Multi-Criteria Decision Making (MCDM) method that addresses decision problems. It highlights the mutual influence of criteria, which is represented both quantitatively and qualitatively. The original use of AHP was to rank a limited number of alternatives for a limited number of criteria.

The problem of selecting Cloud Service can be seen as a MCDM case, where decision makers should select from among a set of alternatives those that best fit their criteria. These criteria are usually conflicting, and each has its importance in the decision making process.

This paper is organized as follows: In Section 2 we introduce the Cloud computing services, Big Data criteria and Cloud Service criteria. Section 3 presents AHP and related work. Section 4 details our proposal based on the application of the AHP in order to select the adequate Cloud Service for Big Data. Then we end up with conclusion and some future works.

## II. Context of Cloud Service and Big Data

### A. Cloud Computing Services

Cloud computing is a model for enabling ubiquitous, convenient, on-demand network access to a shared pool of configurable computing resources (e.g. networks, servers, applications, storage, and services). It can be quickly provisioned and released with minimal management effort or interaction with the service provider. This Cloud model has five main characteristics: (i) On-demand self-service, (ii) Broad network access, (iii) Resource pooling, (iv) Rapid elasticity, and (v) Measured service. It offers three service models: (i) Software as a Service (SaaS), (ii) Platform as a Service (PaaS), and (iii) Infrastructure as a Service (IaaS). And for deployment, it proposes four models : (i) Private Cloud, (ii) Community Cloud, (iii) Public Cloud, and (iv) Hybrid Cloud [1].
Cloud computing offers software dematerialization services [2] as shown in Fig. 1:

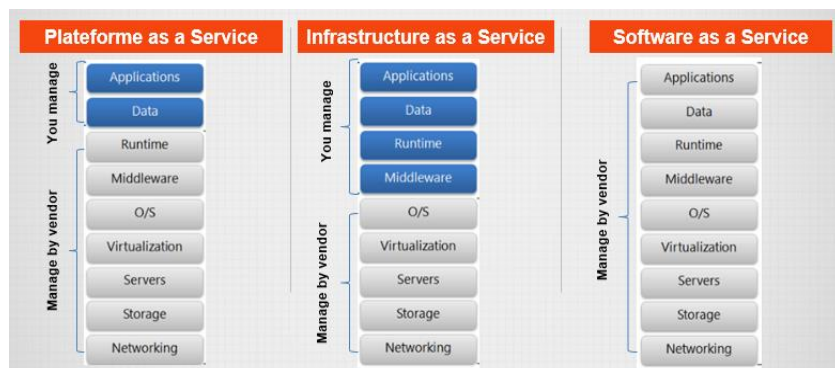

Figure 1. Cloud Services

- PaaS: This service model provides to the consumer the capability to deploy on Cloud infrastructure [3].

- IaaS: This service model provides access to a virtualized IT infrastructure. Virtual machines on which the consumer can install an operating system and applications are made available [4].

- SaaS: In this service model, applications are made available to consumers. Applications can be manipulated using a web browser [5].

### B. Big Data Criteria

Big Data refers to the flood of digital data coming from many digital sources, including sensors, digitizers, scanners, numerical modelling, mobile phones, Internet, videos, e-mails and social network [6]. The main characteristics of Big Data are the five V's: Volume, Velocity, Variety, Value and Veracity [7] as shown in Fig. 2. These Big Data characteristics are the criteria used in our proposed approach.

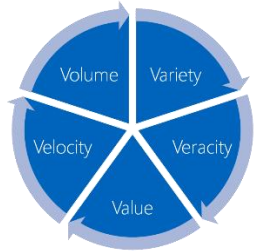

Figure 2. Five V's of Big Data

• Volume (Vol): represents the terabytes of Data produced daily by many platforms such as social networks. This amount of data is definitely diffi-cult to be handled using the existing traditional systems [6].

• Velocity (Vel): Data velocity measures the speed of data creation, streaming, and aggregation. E-Commerce has rapidly increased the speed and rich-ness of data used for different business transactions [7].

• Variety (Var): Data being produced is not of single category as it includes not only the traditional data, but also the semi-structured data from various resources like Web Pages, Web Log Files, social media sites, e-mail, documents, and sensor [6]..

• Value (Val): Is the most important aspect of Big Data. Given that the potential value of Big Data is enormous, it remains useless until its value is discovered. Moreover, since the Big Data's IT infrastructure is costly, its exploitation and value must justify the ROI for investors [8].

• Veracity (Ver): Includes trust and uncertainty issues of the analyzed data and its out coming [9].

*C. Cloud Service Criteria*

Many factors are involved in selection of a Cloud Service. Based on literature reviews, [10] proposed the most common criteria to choose the right Cloud Service namely: Functionality, Vendor Reputation, and Cost. [11] emphasized the importance of the Architecture and Usability, in addition [12] highlighted the Performance factor.

• Functionality (Fun): The functionality is the ability or state of being functional [13]. The need for functionality is obvious; since users will select the right system that provides suitable functions adequate to their needs [14].

• Architecture (Arc): The architecture factors are: (i) The ability to integrate with other applications; (ii) The ability to maintain reasonable response time for users even during peak load; (iii) The ability to remain available for the users for given time windows. It requires vendors to deploy monitoring and diagnostic tools; (iv) Security is considered to be the major concern [11].

• Performance (Per): The performance is determined by the properties of system's constituent parts (e.g. sensors, signal processing, and pattern recognition engine) [13].

• Usability (Usa): Usability includes (i) user interface, (ii) facets such as intuitiveness, (iii) ease-of-use for frequently required tasks and aesthetic nature of graphical elements, (iv) availability of easy-to-use user manuals, (v) eLearning modules, (vi) context-sensitive help, and (vii) offline support that let users work on system in offline mode and let them synchronize once connected to internet [13]. As cited in [14] both usability and functionality are task related and also people related.

• Vendor reputation (Ven): Vendor reputation involves customer perceptions of the vendor's public image, innovativeness, quality of product and service, and commitment to customer satisfaction [14]. Customers can determine vendor reputation based on an evaluation of the vendor's past performance and behavior. Reputation is associated with brand equity and firm credibility; it is also viewed as a sign of trustworthiness [15].

• Cost (Cos): Cost includes one-time implementation's costs and annual subscription. Usually, cost of initial consulting, configuration efforts, etc. is covered under one-time implementation, while cost of hardware and support personnel is covered under annual subscription [13].

The cited criteria in sub section 1 and 2 of the context are the most popular for both Big Data and Cloud services. Such it is difficult to list and use an exhaustive criteria's list; we propose to focus on those already cited. However, our proposed approach is not specified to only those criteria, it could be generalized for several criteria.

### III. ANALYTICAL HIERARCHY PROCESS

#### A. Presentation

AHP is a method for ranking decision alternatives and selecting the best one when the decision maker has multiple criteria. It answers the question, "Which one?". By using AHP, the decision maker selects the best alternative that meets his criteria. He develops a numerical score to rank each alternative based on how well it fits best his needs [16].

The AHP is a powerful, flexible and widely used method for complex problems, which consider the numeric scale for the measurement of quantitative and qualitative performances in a hierarchical structure. This is a value approach to the pairwise comparisons. It is one of the few MCDM approaches capable of handling many criteria. The most important characteristic of the AHP is combining knowledge, experience, and individual opinions in a logical way [17]. AHP was first proposed by Saaty [18], and it is one of the most commonly used methods for solving MCDM problems in different fields [19], such as:

Economic/Management problems [16], Technological problems [20], Political problems [21], Social problems [22], Big Data [17], Cloud Computing [23], [24].

Using AHP, expert opinions and evaluation can be designed in a simple hierarchy system with levels, from the highest to the lowest. The application of AHP to a complex problem involves six essential steps [21]:

• Step1: Define the unstructured problem, and state clearly the objectives and outcomes.

• Step2: Decompose the complex problem into a hierarchical structure with decision elements (criteria and alternatives) as shown in Fig 3.

• Step3: Employ pairwise comparisons among decision elements in order to refill criteria's comparison matrix using the scales presented in Table 1.

• Step4: Repeat step 3 to form the pairwise matrix of alternatives for each criterion and calculate the final vector for each matrix.

• Step5: Check the consistency property of matrices to ensure that the judgments of decision makers are rational.

• Step6: Aggregate the relative weights of decision elements to obtain an overall rating of the alternatives, using the vectors obtained from Step 3 and Step 4.

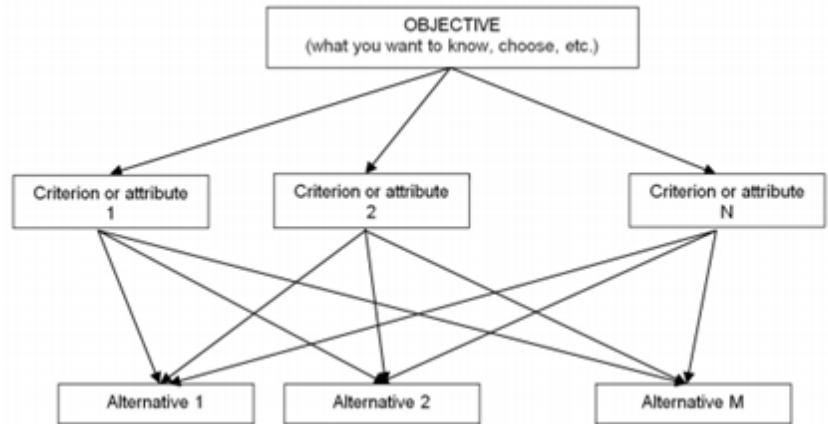

Figure 3. AHP Hierarchy

To make the comparisons cited in the steps above, we need a scale of numbers that indicates how much one element (criterion or alternative) is more important or dominant than another. Table 1 shows the used scale [25].

TABLE 1
PAIRWISE COMPARISON SCALE

| Intensity of importance | Definition | Explanation |
|---|---|---|
| 1 | Equal importance | Two activities contribute equally to the objective |
| 2 | Weak or slight | |
| 3 | Moderate importance | Experience and judgement slighty favour one activity over another |
| 4 | Moderate plus | |
| 5 | Strong importance | Experience and judgement slighty favour one activity over another |
| 6 | Strong plus | |
| 7 | Very strong or demonstrated importance | An activity is favoured one activity over another; its dominance demonstrated in practice |
| 8 | Very, very strong | |
| 9 | Extreme importance | The evidence favouring one activity over another is of the highest possible order of affirmation |

B. Related Work

AHP has been used in many works dealing with the selection of Cloud Service. [26] presented a consumer-centered Cloud Service selection based on AHP. They adopted five criteria, namely: (i) response time, (ii) throughput, (iii) availability, (iv) reliability, and (v) cost. Weights are assigned to each one of these criteria to exhibit there importance in the decision making process. To validate their proposals, they performed tests on three types of Cloud services.

Work presented in [27] concerns an AHP-based methodology for selecting a Cloud Service for companies that would like to reduce the cost of using the software. The main contribution of this work is defining other costs beside the financial one; such as costs related to scalability and risk. This is done by calculating the gain of scalability, and this by improving the agility obtained through Cloud Services. In [28], authors presented a mathematical decision model for selecting Cloud Services based on Linear Optimization and using AHP.

[29] proposed an AHP-based framework for QoS assurance to provide Cloud Services that meets both user and application requirements in terms of Service Level Agreements. While, [11] suggested an AHP-based approach for the selection of SaaS Cloud Services. [27] Presented a Cloud Service evaluation index system based on AHP. They used four evaluation criteria: (i) cost, (ii) security, (iii) reputation, and (iv) QoS. AHP is used to determine the weights of each criteria and to score nominated Cloud Services

[17] Provide an overview of Big Data analytics platforms. They proposed an AHP model, which offers a significant evaluation method that help private and public institutions selecting the suitable Big Data analytics platforms.

To summarize, we can point out that AHP is a tool that has found its use in many areas, especially in the case of Cloud Service selection. The success of the method is a consequence of its simplicity and robustness. Despite all the work cited below based on AHP, there is a lack of its use for selecting the right Cloud Service for Big Data processing.

## IV. Our Proposal Approach

Based on both robustness and simplicity of the AHP, we conducted our works to select the Cloud Service adequate to Big Data. We applied this approach on a case study in order to validate its efficity. The studied case is about the selection of the adequate Cloud Service for Big Data in the context of National Health Service (NHS) of United Kingdom (UK).

NHS is the organization responsible for all healthcare services of UK. Its main concern is to manage the entire medical field characterized by massive, critical and heterogeneous Data. In recent years, the volume of data available within the NHS has increased exponentially. Modern day computing power, combined with the drop in price of data storage, means Big Data analytics is becoming more achievable [30].

To deal with the particularity of medical Data, NHS has a great potential to achieve more using the Cloud Services [31] to store and analyze their Data. This will induct a reduction in storage costs and it offers an open pathway to Big Data analytics.

However, NHS is facing the problem of moving its data in the Cloud environment. To deal with this issue, we propose to apply our approach to assist NHS select the most adequate Cloud Service.

First, we start with hypothesis to apply the AHP to our case study. We suppose that:

1.    We focused on three kinds of Cloud Service models (SAAS, PASS, and IAAS).

2.    We based on criteria mentioned in section 2: functionality, usability, architecture, vendor, performance, cost, volume, velocity, variety, value and veracity

TABLE 2
BIG DATA CHARACTERISTICS OF NHS [32]

| 5Vs of Big Data | NHS's Data Characteristics |
|---|---|
| Volume | Increasing size |
| Velocity | Speed of generation and processing |
| Variety | Heterogeneous Formats (XML, CSV, Multimedia, etc.) |
| Veracity | Critical authenticity |
| Value | Significant |

As mentioned in AHP's section, we have 6 steps to be applied to the studied case:

Step1: Our issue is to select a Cloud Service adequate to NHS's Big Data. In or-der to achieve this goal, multiple criteria to select a Cloud Service were determined in Section 2 and then compared according to their importance. Finally, the most appropriate Cloud Service will be selected according to the predetermined criteria.

Step2: As it is shown in Fig. 4, the hierarchy structure of our announced problem is composed from 3 layers. First layer is the one that defines the main goal of the problem: "appropriate Cloud Service for NHS's Big Data". Layer 2 defines the selection's criterion. Layer 3 covers alternatives of the Cloud Services (SAAS, PAAS, IAAS).

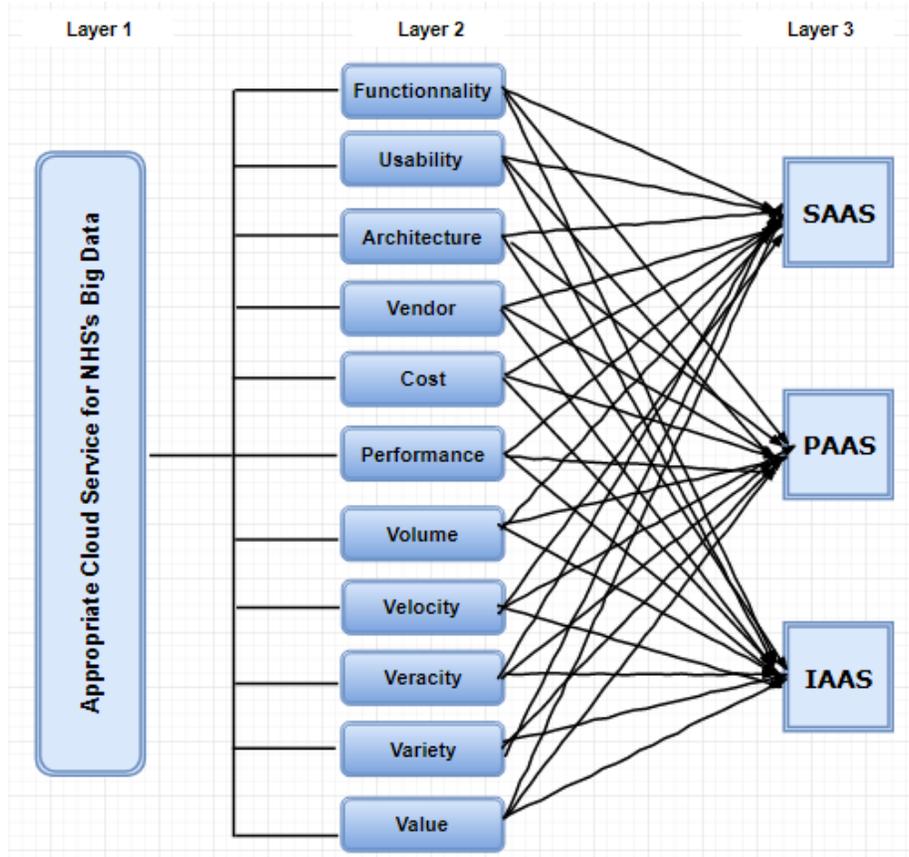

Figure 4. Hierarchy structure of AHP's step 2

Step3: We have constructed a pairwise comparison matrix, as shown in Table 3. This matrix is composed of 2 parts: a first part (in blue) concerns the comparison of the criteria and the second (in green) included the calculated results based on data of the first part of the matrix .

TABLE 3
Pairwise Matrix

| | Fun | Usa | Arc | Ven | Cos | Per | Vol | Vel | Var | Val | Ver | %Fun | % Usa | %Arc | % Ven | % Cos | % Per | % Vol | % Vel | % Var | % Val | % Ver | Sum%VCRI | Sum%VCRI/N |
|---|---|---|---|---|---|---|---|---|---|---|---|---|---|---|---|---|---|---|---|---|---|---|---|---|
| Fun | 1 | 5 | 7 | 3 | 3 | 5 | 1 | 5 | 3 | 3 | 3 | 0,227 | 0,138 | 0,632 | 0,077 | 0,109 | 0,261 | 0,207 | 0,322 | 0,2217 | 0,083 | 0,093 | 2,371 | 0,216 |
| Usa | 1/5 | 1 | 1/7 | 5 | 3 | 1/7 | 1/7 | 1/5 | 3 | 3 | 3 | 0,045 | 0,028 | 0,013 | 0,128 | 0,109 | 0,007 | 0,009 | 0,012 | 0,0246 | 0,083 | 0,093 | 0,572 | 0,052 |
| Arc | 1/7 | 7 | 1 | 5 | 3 | 5 | 1 | 3 | 3 | 3 | 5 | 0,032 | 0,193 | 0,090 | 0,128 | 0,109 | 0,261 | 0,207 | 0,193 | 0,2217 | 0,138 | 0,093 | 1,668 | 0,152 |
| Ven | 1/3 | 1/5 | 1/5 | 1 | 1 | 1/3 | 1/7 | 1/5 | 1/3 | 1/3 | 1/3 | 0,076 | 0,006 | 0,018 | 0,026 | 0,036 | 0,017 | 0,009 | 0,013 | 0,0246 | 0,009 | 0,010 | 0,255 | 0,024 |
| Cos | 1/3 | 1/3 | 1/3 | 1 | 1 | 1/3 | 1/7 | 1/3 | 1/3 | 3 | 3 | 0,076 | 0,009 | 0,030 | 0,026 | 0,036 | 0,017 | 0,030 | 0,021 | 0,0246 | 0,138 | 0,155 | 0,562 | 0,051 |
| Per | 1/5 | 7 | 1/5 | 3 | 3 | 1 | 1/5 | 1 | 1 | 7 | 7 | 0,045 | 0,193 | 0,018 | 0,077 | 0,109 | 0,052 | 0,041 | 0,064 | 0,0739 | 0,193 | 0,216 | 1,084 | 0,099 |
| Vol | 1 | 7 | 1 | 7 | 7 | 5 | 1 | 3 | 3 | 5 | 3 | 0,227 | 0,193 | 0,090 | 0,179 | 0,255 | 0,261 | 0,207 | 0,193 | 0,2217 | 0,138 | 0,093 | 2,069 | 0,187 |
| Vel | 1/5 | 3 | 1/3 | 5 | 3 | 1 | 1/3 | 1 | 3 | 3 | 3 | 0,045 | 0,083 | 0,030 | 0,128 | 0,109 | 0,052 | 0,069 | 0,064 | 0,0739 | 0,083 | 0,093 | 0,831 | 0,076 |
| Var | 1/3 | 1/3 | 1/3 | 3 | 3 | 1 | 1/3 | 1/3 | 1 | 3 | 3 | 0,076 | 0,009 | 0,030 | 0,077 | 0,109 | 0,052 | 0,069 | 0,064 | 0,0739 | 0,083 | 0,093 | 0,865 | 0,079 |
| Val | 1/3 | 1/3 | 1/5 | 3 | 1/5 | 1/7 | 1/5 | 1/3 | 1/3 | 1 | 1 | 0,076 | 0,009 | 0,018 | 0,077 | 0,007 | 0,007 | 0,041 | 0,021 | 0,0246 | 0,028 | 0,031 | 0,341 | 0,031 |
| Ver | 1/3 | 1/3 | 1/5 | 3 | 1/5 | 1/7 | 1/3 | 1/3 | 1/3 | 1 | 1 | 0,076 | 0,009 | 0,030 | 0,077 | 0,007 | 0,007 | 0,069 | 0,021 | 0,0246 | 0,028 | 0,031 | 0,380 | 0,035 |
| SumVCRI | 4,410 | 36,200 | 11,076 | 39,000 | 27,400 | 19,095 | 4,829 | 15,533 | 13,533 | 36,333 | 32,333 | 1,000 | 1,000 | 1,000 | 1,000 | 1,000 | 1,000 | 1,000 | 1,0000 | 1,0000 | 1,000 | 1,000 | 11,000 | 1,000 |

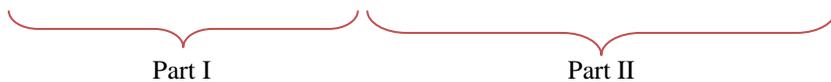

Part I          Part II

The criteria are compared with each other while respecting the initial objective. We have assigned a value between 1 and 9 as specified in Table 1, this value indicates the degree of importance of one criterion with respect to the other based on the needs of the NHS [32] Table 4 shows an example of choosing some value when comparing two criteria. We used these values to fill the part I of the matrix in Table 3.

TABLE 4
Example of pairwise comparison scales for the criterion Functionality

| Intensity of importance | Definition | Explanation |
|---|---|---|
| 1 | Functionality/ volume | Functionality and volume contribute equally to the goal |
| 5 | Functionality/velocity | Functionality has a strong importance than velocity |
| 1/5 | Velocity/functionality | Inverse of functionaliy/velocity |

For the part II of the matrix, we set CRI the criteria: $CRI_i \in$ {Fun, Usa, Arc, Ven, Cos, Per, Vol, Vel, Var, Val, Ver} and $VCRI_{ij}$ the pairwise comparison value of the criterion i and the criterion j, where $i,j \in$ {1,2,....,N}  N number of criteria. We Set $SumVCRI_j$ as the sum of column j's elements.

$$SumVCRI_j = \sum_{i=1}^{N} VCRI_{ij}. \qquad (1)$$

$\%VCRI_{ij}$ represent the percentage of each pairwise comparison value $VCRI_{ij}$.

$$\%VCRI_{ij} = VCRI_{ij} / SumVCRI_j \qquad (2)$$

$Sum\%VCRI_i$ is the sum of all the $\%VCRI_{ij}$ of the line i.

$$Sum\%VCRI_i = \sum_{j=1}^{N} \%VCRI_{ij} \qquad (3)$$

Finally, Sum%VCRIi must be divided by N for each line i in order to obtain the fi-nal vector to be used in step 6.

Step 4: According to NHS characteristics (Connecting for Health,2011) and for every criterion CRIi we repeat step 3 to form the pairwise matrix of alternatives ALT for each criterion CRI and calculate the final vector for each matrix as shown in Table 5 to 15. We set:

1) ALT the alternatives: ALTi ϵ {SAAS, PAAS, and IAAS}

2) VALTij the pairwise comparison value of the alternative i and the alternative j

3) SumVALTj as the sum of column j's elements.

i,j ϵ {1,2,….,M}  M number of alternatives.

$$\text{SumVALT}_j = \sum_{i=1}^{M} \text{VALTij} \qquad (4)$$

%VALTij represent the percentage of each pairwise comparaison value VALTij.

$$\%\text{VALT}_{ij} = \text{VALTij} / \text{SumVALTj} \qquad (5)$$

Sum%VALT$_i$ is the sum of all the %ALT$_{ij}$ of the line i.

$$\text{Sum}\%\text{VALT}_i = \sum_{j=1}^{M} \%\text{VALTij} \qquad (6)$$

TABLE 5
Pairwise comparison of alternatives for functionality

| Fun | SAAS | PAAS | IAAS | % SAAS | % PAAS | %IAAS | Sum%VALT | Sum%VALT/M |
|-----|------|------|------|--------|--------|-------|----------|------------|
| SAAS | 1 | 5 | 7 | 0,745 | 0,714 | 0,778 | 2,237 | 0,847 |
| PAAS | 1/5 | 1 | 1 | 0,149 | 0,143 | 0,111 | 0,403 | 0,153 |
| IAAS | 1/7 | 1 | 1 | 0,106 | 0,143 | 0,111 | 0,360 | 0,137 |
| SumVALT | 1,343 | 7,000 | 9,000 | 1,000 | 1,000 | 0,889 | 2,640 | 1,000 |

TABLE 6
Pairwise comparison of alternatives for usability

| Usa | SAAS | PAAS | IAAS | % SAAS | % PAAS | %IAAS | Sum%VALT | Sum%VALT/M |
|-----|------|------|------|--------|--------|-------|----------|------------|
| SAAS | 1 | 5 | 7 | 0,745 | 0,789 | 0,636 | 2,171 | 0,724 |
| PAAS | 1/5 | 1 | 3 | 0,149 | 0,158 | 0,273 | 0,580 | 0,193 |
| IAAS | 1/7 | 1/3 | 1 | 0,106 | 0,053 | 0,091 | 0,250 | 0,083 |
| SumVALT | 1,343 | 6,333 | 11,000 | 1,000 | 1,000 | 1,000 | 3,000 | 1,000 |

TABLE 7
Pairwise comparison of alternatives for vendor reputation

| Ven | SAAS | PAAS | IAAS | % SAAS | % PAAS | %IAAS | Sum%VALT | Sum%VALT/M |
|-----|------|------|------|--------|--------|-------|----------|------------|
| SAAS | 1 | 1 | 1 | 0,333 | 0,333 | 0,333 | 1,000 | 0,333 |
| PAAS | 1 | 1 | 1 | 0,333 | 0,333 | 0,333 | 1,000 | 0,333 |
| IAAS | 1 | 1 | 1 | 0,333 | 0,333 | 0,333 | 1,000 | 0,333 |
| SumVALT | 3,000 | 3,000 | 3,000 | 1,000 | 1,000 | 1,000 | 3,000 | 1,000 |

TABLE 8
Pairwise comparison of alternatives for architecture

| Arc | SAAS | PAAS | IAAS | % SAAS | % PAAS | %IAAS | Sum%VALT | Sum%VALT/M |
|-----|------|------|------|--------|--------|-------|----------|------------|
| SAAS | 1 | 1/3 | 1/3 | 0,143 | 0,077 | 0,200 | 0,420 | 0,140 |
| PAAS | 3 | 1 | 1/3 | 0,429 | 0,231 | 0,200 | 0,859 | 0,286 |
| IAAS | 3 | 3 | 1 | 0,429 | 0,692 | 0,600 | 1,721 | 0,574 |
| SumVALT | 7,000 | 4,333 | 1,667 | 1,000 | 1,000 | 1,000 | 3,000 | 1,000 |

TABLE 9
Pairwise comparison of alternatives for cost

| Cos | SAAS | PAAS | IAAS | % SAAS | % PAAS | %IAAS | Sum%VALT | Sum%VALT/M |
|---|---|---|---|---|---|---|---|---|
| SAAS | 1 | 5 | 7 | 0,745 | 0,789 | 0,636 | 2,171 | 0,724 |
| PAAS | 1/5 | 1 | 3 | 0,149 | 0,158 | 0,273 | 0,580 | 0,193 |
| IAAS | 1/7 | 1/3 | 1 | 0,106 | 0,053 | 0,091 | 0,250 | 0,083 |
| SumVALT | 1,343 | 6,333 | 11,000 | 1,000 | 1,000 | 1,000 | 3,000 | 1,000 |

TABLE 10
Pairwise comparison of alternatives for performance

| Per | SAAS | PAAS | IAAS | % SAAS | % PAAS | %IAAS | Sum%VALT | Sum%VALT/M |
|---|---|---|---|---|---|---|---|---|
| SAAS | 1 | 5 | 3 | 0,652 | 0,714 | 0,600 | 1,966 | 0,655 |
| PAAS | 1/5 | 1 | 1 | 0,130 | 0,143 | 0,200 | 0,473 | 0,158 |
| IAAS | 1/3 | 1 | 1 | 0,217 | 0,143 | 0,200 | 0,560 | 0,187 |
| SumVALT | 1,533 | 7,000 | 5,000 | 1,000 | 1,000 | 1,000 | 3,000 | 1,000 |

TABLE 11
Pairwise comparison of alternatives for volume

| Vol | SAAS | PAAS | IAAS | % SAAS | % PAAS | %IAAS | Sum%VALT | Sum%VALT/M |
|---|---|---|---|---|---|---|---|---|
| SAAS | 1 | 2 | 2 | 0,500 | 0,571 | 0,400 | 1,471 | 0,490 |
| PAAS | 1/2 | 1 | 2 | 0,250 | 0,286 | 0,400 | 0,936 | 0,312 |
| IAAS | 1/2 | 1/2 | 1 | 0,250 | 0,143 | 0,200 | 0,593 | 0,198 |
| SumVALT | 2,000 | 3,500 | 5,000 | 1,000 | 1,000 | 1,000 | 3,000 | 1,000 |

TABLE 12
Pairwise comparison of alternatives for velocity

| Vel | SAAS | PAAS | IAAS | % SAAS | % PAAS | %IAAS | Sum%VALT | Sum%VALT/M |
|---|---|---|---|---|---|---|---|---|
| SAAS | 1 | 3 | 3 | 0,600 | 0,600 | 0,600 | 1,800 | 0,600 |
| PAAS | 1/3 | 1 | 1 | 0,200 | 0,200 | 0,200 | 0,600 | 0,200 |
| IAAS | 1/3 | 1 | 1 | 0,200 | 0,200 | 0,200 | 0,600 | 0,200 |
| SumVALT | 1,667 | 5,000 | 5,000 | 1,000 | 1,000 | 1,000 | 3,000 | 1,000 |

TABLE 13
Pairwise comparison of alternatives for veracity

| Ver | SAAS | PAAS | IAAS | % SAAS | % PAAS | %IAAS | Sum%VALT | Sum%VALT/M |
|---|---|---|---|---|---|---|---|---|
| SAAS | 1 | 5 | 3 | 0,652 | 0,789 | 0,429 | 1,870 | 0,623 |
| PAAS | 1/5 | 1 | 3 | 0,130 | 0,158 | 0,429 | 0,717 | 0,239 |
| IAAS | 1/3 | 1/3 | 1 | 0,217 | 0,053 | 0,143 | 0,413 | 0,138 |
| SumVALT | 1,533 | 6,333 | 7,000 | 1,000 | 1,000 | 1,000 | 3,000 | 1,000 |

TABLE 14
Pairwise comparison of alternatives for variety

| Var | SAAS | PAAS | IAAS | % SAAS | % PAAS | %IAAS | Sum%VALT | Sum%VALT/M |
|---|---|---|---|---|---|---|---|---|
| SAAS | 1 | 1 | 1 | 0,333 | 0,333 | 0,333 | 1,000 | 0,333 |
| PAAS | 1 | 1 | 1 | 0,333 | 0,333 | 0,333 | 1,000 | 0,333 |
| IAAS | 1 | 1 | 1 | 0,333 | 0,333 | 0,333 | 1,000 | 0,333 |
| SumVALT | 3,000 | 3,000 | 3,000 | 1,000 | 1,000 | 1,000 | 3,000 | 1,000 |

TABLE 15
Pairwise comparison of alternatives for value

| Val | SAAS | PAAS | IAAS | % SAAS | % PAAS | %IAAS | Sum%VALT | Sum%VALT/M |
|---|---|---|---|---|---|---|---|---|
| SAAS | 1 | 5 | 3 | 0,652 | 0,789 | 0,429 | 1,870 | 0,623 |
| PAAS | 1/5 | 1 | 3 | 0,130 | 0,158 | 0,429 | 0,717 | 0,239 |
| IAAS | 1/3 | 1/3 | 1 | 0,217 | 0,053 | 0,143 | 0,413 | 0,138 |
| SumVALT | 1,533 | 6,333 | 7,000 | 1,000 | 1,000 | 1,000 | 3,000 | 1,000 |

In Table 5 to15, Sum%VALT must be divided by M for each line i in order to obtain the final vector to be used in step 6 . It represent the averages of each alternative ALT for each criteria CRI.

Step5: we checked the consistency property of matrices, in fact as it shown in previous tables the obtained results are homogeneous.

Step 6: We used vector Sum%VCRI/N obtained in step 3 as shown in Table 4 to fill the first line of Table 16

For each criterion CRI we used the vectors Sum%VALT/M obtained in step 4 as shown in Tables 5 to15 in order to fill columns 1 to N in Table 16.

we calculate Result according to the following formula.

$$\text{Result}_i = \sum_{j=1}^{N} \frac{\text{Sum\%VCRI}_i}{N} * \frac{\text{Sum\%VALT}_i}{M} \qquad (7)$$

TABLE 16
Synthesis results

| | Fun | Usa | Arc | Ven | Cos | Per | Vol | Vel | Var | Val | Ver | Result |
|---|---|---|---|---|---|---|---|---|---|---|---|---|
| Sum%VCRI/N | 0,216 | 0,052 | 0,152 | 0,024 | 0,051 | 0,099 | 0,191 | 0,076 | 0,079 | 0,031 | 0,035 | |
| Sum%VALT/M (SAAS) | 0,847 | 0,724 | 0,140 | 0,333 | 0,724 | 0,655 | 0,490 | 0,600 | 0,333 | 0,623 | 0,623 | 0,557 |
| Sum%VALT/M (PAAS) | 0,153 | 0,193 | 0,286 | 0,333 | 0,193 | 0,158 | 0,312 | 0,200 | 0,333 | 0,239 | 0,239 | 0,236 |
| Sum%VALT/M (IAAS) | 0,137 | 0,083 | 0,574 | 0,333 | 0,083 | 0,187 | 0,198 | 0,200 | 0,333 | 0,138 | 0,138 | 0,240 |

Result and synthesis:
Following the AHP methodology, paired comparisons of criteria, and the com-parison for alternatives by criterion was made according to their intensity of im-portance and respecting a fundamental scale of absolute numbers. Then, all the calculations were performed to find normalized values for each alternative. The final decision is taken based on these normalized values.

Table 16 shows the final weights for the selected alternatives for NHS use case. Based on requirement defined in table 2, the normalized values obtained for each alternative are: 0.557 for SAAS, 0.240 for IAAS and 0.236 for PAAS. We can summarize that SAAS is the most qualified Cloud Service for NHS case.

## VI. Conclusion and Perspectives

In this paper, we provide an overview of the use of AHP in the field of cloud service, and Big Data platforms. We highlight the use of the method for selecting an appropriate cloud service for Big Data. This overview allowed us to deduce that there is a lack of AHP's use in this area.

We developed an approach that offers a simple and efficient evaluation method that can help organizations to select the most suitable cloud service ade-quate to their Big Data. To address this issue we based on AHP. In order to pro-vide complete understanding of the process we apply our approach to NHS case study.

The selection of best cloud service satisfying most of the requirements among the available alternatives is a MCDM problem. The AHP process involves multi-ple criteria and alternatives. We begin by developing a hierarchy model of 3 lay-ers: (i) goal, (ii) criteria, and (iii) alternatives. We first compare criteria pairwise with respect to the desired goal. Then we compare the alternatives pairwise for each criterion.

We calculate the relative weights of decision elements to obtain synthesis results. Finaly, we can make a final decision based on these synthesis results.

In our future works, we intend to automate our approach in order to facilitate its use, reduce calculations rate and obtain the result in few times.

The weight assignement strategy will be difficult in the case where number of cri-teria is huge. To overcome this difficulty, we will use fuzzy logic to affect weights.

We will conduct a comparative study of different Cloud Service selection meth-ods to determine the strengths and weaknesses of each.

We also plan to treat the dependability of Big Data in cloud environment in order to determine the attitude of Cloud systems to complete the features required by Big Data.